# Một mô hình không gian vectơ tổng quát cho truy hồi thông tin dựa trên ontology

## A Generalized Vector Space Model for Ontology-Based Information Retrieval

Ngô Minh Vương và Cao Hoàng Trụ

Khoa Khoa Học và Kỹ Thuật Máy Tính

Trường Đại Học Bách Khoa TpHCM, Việt Nam

{vuong.cs@gmail.com}

***Abstract***: *Named entities (NE) are objects that are referred to by names such as people, organizations and locations. Named entities and keywords are important to the meaning of a document. We propose a generalized vector space model that combines named entities and keywords. In the model, we take into account different ontological features of named entities, namely, aliases, classes and identifiers. Moreover, we use entity classes to represent the latent information of interrogative words in Wh-queries, which are ignored in traditional keyword-based searching. We have implemented and tested the proposed model on a TREC dataset, as presented and discussed in the paper.*

***Tóm tắt***: *Thực thể có tên (NE) là những đối tượng được tham khảo đến bằng tên như con người, tổ chức, nơi chốn. Các thực thể và từ khóa có ý nghĩa quan trọng đối với một tài liệu. Chúng tôi đề xuất một mô hình không gian vectơ tổng quát kết hợp thực thể có tên và từ khóa. Trong mô hình này, chúng tôi xét đến các đặc điểm ontology khác nhau của thực thể có tên như bí danh, lớp và định danh. Hơn nữa, chúng tôi dùng lớp thực thể để biểu diễn thông tin tiềm ẩn của các từ để hỏi trong các truy vấn Wh, bị bỏ qua trong tìm kiếm dựa trên từ khóa truyền thống. Chúng tôi đã hiện thực và kiểm tra mô hình đề xuất trên một tập dữ liệu TREC như được trình bày và thảo luận trong bài báo.*

***Các từ khóa***: *truy hồi thông tin, mô hình không gian vectơ, ontology, thực thể có tên, từ khóa.*

## I. GIỚI THIỆU

Như chúng ta đã biết, ngày nay WWW lưu trữ mọi tài nguyên thông tin chia sẻ trên toàn cầu, và các thông tin này đang được phát triển nhanh chóng. Nhu cầu khai thác thông tin trên WWW là rất lớn. Theo bản báo cáo tháng 05/2007 của ComScore[1], có khoảng 772 triệu người trên 15 tuổi sử dụng Internet với tần suất thường xuyên. Vấn đề đặt ra là làm thế nào để có thể khai thác và sử dụng tài nguyên đó một cách hiệu quả.

Dữ liệu được lưu trữ trên Web là bán cấu trúc. Đa phần các thông tin này được thể hiện dưới dạng ngôn ngữ tự nhiên, chỉ phù hợp cho con người đọc hiểu, và các hệ thống truy hồi thông tin hiện nay sử dụng từ khóa để tìm kiếm các tài liệu. Nhiều ngữ nghĩa trong tài liệu hoặc yêu cầu của người dùng bị mất khi ta biểu diễn văn bản của nó thành một tập các từ khóa. Việc so trùng giữa một tài liệu và yêu cầu của người dùng là sự so trùng giữa tập các từ khóa đại diện cho tài liệu và tập các từ khóa đại diện cho yêu cầu. Vì vậy, có nhiều tài liệu được trả về theo yêu cầu của người sử dụng thông qua việc so trùng các từ khóa là không thực sự phù hợp với yêu cầu này.

Để khắc phục nhược điểm của mô hình truy hồi thông tin theo từ khóa, cùng với sự ra đời và phát triển của Web ngữ nghĩa (Semantic Web [9]), ontology và cơ sở tri thức (Knowledge Base – KB) bắt đầu được sử dụng rộng rãi trong các nghiên cứu phát triển hệ thống truy hồi thông tin có ngữ nghĩa ([5], [15], [12], [7], [14]). Thực thể có tên (named entity, NE) là đối tượng được đặc tả trong ontology và KB. NE được sử dụng rộng rãi trong các ứng dụng rút trích thông tin, trả lời các câu hỏi, xử lý ngôn ngữ tự nhiên, được đề cập đến lần đầu tại Message Understanding Conference (MUC) những năm 1990 ([10]). NE là con người, tổ chức, nơi chốn, và những đối tượng khác được tham khảo bằng tên. Việc xử lý các từ khóa chỉ đòi hỏi về sự xác định ngữ nghĩa từ vựng thông thường còn việc xử lý các NE cần đến tri thức về đối tượng của thế giới đang xét

Trong quá trình đọc hiểu nghĩa của văn bản, cấp độ đầu tiên là hiểu nghĩa của từ. Đây là cấp độ xử lý theo từ khóa. Cấp độ cao hơn là nhận ra các đối tượng được nói đến trong văn bản, được xác định bởi NE và các mối quan hệ của chúng. Do đó, các NE được đề cập đến trong tài liệu có một vị trí quan trọng về mặt ngữ nghĩa của tài liệu đó. Vì thế các tài liệu Web có chú thích ngữ nghĩa cho NE sẽ giúp cho việc truy hồi và khai thác thông tin trên đó được chính xác và hiệu quả hơn. Các nghiên cứu [11], [6], [8], [5], [13], [3] đã khai thác NE nhằm tăng hiệu suất truy hồi thông tin, như sẽ được phân tích và so sánh ở Phần V với mô hình chúng tôi đề xuất trong bài báo này.

---

[1] www.comscore.com/press/

NE là thành phần quan trọng trong tài liệu. Tuy nhiên NE chưa thể hiện tất cả nội dung của tài liệu và trong nhiều ngữ cảnh thì từ khóa giữ vai trò quan trọng về ngữ nghĩa của một tài liệu. Ví dụ như trong trường hợp, người dùng mong muốn tìm kiếm các tài liệu nói về: "sự ô nhiễm ở thành phố" thì NE *thành phố* và từ khóa *sự ô nhiễm* đều có vai trò quan trọng trong việc tìm kiếm các tài liệu phù hợp. Do đó chúng tôi đề xuất việc nghiên cứu mô hình truy hồi thông tin kết hợp NE và từ khóa.

Công trình [3] đã đề xuất một mô hình không gian vectơ hợp nhất NE và từ khóa, tức là mỗi phần tử của không gian vectơ có thể là NE hoặc có thể là từ khóa. Trong đó, có khai thác đầy đủ tên, lớp, định danh của NE. Thêm vào đó, [3] dựa vào đặc tính của NE để tiến hành mở rộng vectơ tài liệu mà không làm thay đổi nội dung ban đầu của tài liệu. Tuy nhiên, công trình [3] còn các hạn chế là chưa xử lý các từ để hỏi trong các truy vấn *Wh* và chưa xử lý các bí danh (alias), tức là tên khác, của các NE một cách hiệu quả về bộ nhớ và thời gian xử lý. Bài báo này khắc phục các hạn chế nêu trên và tiến hành thí nghiệm hệ thống trên một tập kiểm tra lớn hơn, của TREC gồm 130.000 tài liệu và 124 truy vấn.

## II. TRUY HỒI THÔNG TIN THEO THỰC THỂ CÓ TÊN VÀ TỪ KHÓA

### 1. Thực thể có tên

Một NE có thể có nhiều tên khác nhau, được gọi là các bí danh của NE đó. Mặt khác, cũng có nhiều NE khác nhau nhưng có cùng tên. Chúng tôi biểu diễn mỗi NE bằng bộ ba (*tên, lớp, định danh*).

Ví dụ, xem xét đoạn văn bản sau từ trang Web của UBND TP.HCM [2] được viết vào ngày 11-11-2006:

"*Chiều 10-11-2006, tại trụ sở UBND TP.HCM, tập đoàn Intel đã công bố quyết định mở rộng diện tích nhà máy lắp ráp và kiểm định chíp bán dẫn tại Khu Công Nghệ Cao Sài Gòn từ 13.900 m² lên 46.450 m² đồng thời tăng tổng vốn đầu tư từ 300 triệu USD lên 1 tỷ USD.*

*[…]*

*Ông Brian M.Krzanich, Phó Chủ tịch, Tổng giám đốc Lắp ráp và Kiểm tra sản phẩm của Tập đoàn Intel cho biết nhà máy được xây dựng tại Việt Nam lần này sẽ là nhà máy đầu tiên có quy mô lớn nhất trong hệ thống lắp ráp và kiểm định của Intel.*"

Ở đây, *UBND TP.HCM*, *Intel*, *Khu Công Nghệ Cao Sài Gòn*, *Brian M.Krzanich* và *Việt Nam* là các NE.

Mỗi NE có thể được chú thích đầy đủ bởi tên, lớp, định danh nếu nó tồn tại trong cơ sở tri thức. Ví dụ, một chú thích đầy đủ của Intel là bộ ba (*Intel/ Company/Company_123*), với *Company_123* là định danh của NE Intel. Vì sự nhập nhằng trong một ngữ cảnh hoặc trong sự thực hiện của một phương thức nhận diện, một NE có thể không được chú thích đầy đủ hoặc có thể có nhiều sự chú thích. Ví dụ như *Brian M.Krzanich* có thể được nhận diện như một người, mặc dù nó không tồn tại trong cơ sở tri thức và định danh của nó không được xác định. Trong khi đó, *Khu Công Nghệ Cao Sài Gòn* có thể là khu công nghiệp (thuộc lớp vị trí) hoặc là một ban quản lý (thuộc lớp tổ chức).

Tóm lại, sự chú thích của một NE có thể là một trong các dạng sau:

1. Chỉ có tên: khi không nhận diện được lớp của nó
2. Chỉ có tên và lớp: khi không nhận diện được định danh.
3. Đầy đủ tên, lớp và định danh.

Chúng tôi lưu ý là các tên và lớp của một NE có thể được dẫn xuất từ định danh của nó dựa trên ontology và cơ sở tri thức. Hơn nữa, với một chú thích của một NE cho trước, có thể dẫn xuất ra các lớp cha và các tên khác của NE này.

### 2. Mô hình không gian viectơ kết hợp thực thể có tên và từ khóa

Mô hình không gian vectơ (VSM) là một mô hình truy hồi thông tin có nhiều ưu điểm và được sử dụng rộng rãi. Với các hạn chế chung của các mô hình truy hồi thông tin theo từ khóa, mô hình VSM dựa trên từ khóa không thể thỏa mãn người dùng với các dạng truy vấn như:

1. Tìm tài liệu về tổ chức thương mại.
2. Tìm tài liệu về Sài Gòn.
3. Tìm tài liệu về thành phố Paris.
4. Tìm tài liệu về thành phố Paris thuộc bang Texas của nước Mỹ.

Đó là bởi vì, ở truy vấn thứ nhất, một tài liệu kết quả đúng không nhất thiết phải chứa từ khóa *tổ chức thương mại*, nhưng chỉ cần chứa một hoặc một vài NE có lớp là *Tổ Chức Thương Mại*, như là NIKE, SONY,... . Ở truy vấn thứ hai, các tài liệu đề cập đến thành phố *Sài Gòn* bởi một số tên khác của nó như là "thành phố Hồ Chí Minh" hoặc "TPHCM" cũng là các kết quả đúng. Mặc khác, các tài liệu kết quả có thể chứa các NE khác có tên là *Sài Gòn*, ví dụ như "sông Sài Gòn" hay "ngân hàng Sài Gòn". Ở truy vấn thứ ba, người sử dụng không mong chờ nhận các tài liệu về *Paris* như là một trường đại học hoặc một ngân hàng. Ở đây, chúng tôi lưu ý là có nhiều thành phố tên *Paris* trên thế giới, và bất kỳ bài liệu nào đề cập đến một trong các thành phố này đều là kết quả đúng. Truy vấn thứ tư nhắm tới các tài liệu chứa NE có định danh, được xác định hoàn toàn.

Tuy nhiên, ở trong nhiều trường hợp, NE không đại diện được đầy đủ nội dung của tài liệu hoặc truy vấn. Ví dụ, trong truy vấn về "*biểu tình ở Thái Lan*", nếu chỉ xem xét về thực thể có tên, ta chỉ có được thông tin là nước *Thái Lan*, và bỏ qua thông tin quan trọng của từ khóa *biểu tình*. Bên cạnh đó, có những truy vấn không có NE. Do đó, các mô hình không gian vectơ dựa trên NE cần phải được cải tiến, kết hợp với từ khóa để biểu diễn được tốt hơn ngữ nghĩa của tài liệu và truy vấn, từ đó nâng cao hiệu quả truy hồi.

---

[2] www.hochiminhcity.gov.vn

Công trình [3] đề xuất mô hình kết hợp NE và từ khóa, có khảo sát đầy đủ ba đặc trưng của NE. Kiến trúc của hệ thống truy hồi thông tin theo NE và từ khóa được thể hiện ở Hình 1. Nó chứa một ontology và một cơ sở tri thức. Môđun *Nhận diện và chú thích NE* được dùng để rút trích và nhúng thông tin về NE vào trong tài liệu thô, trước khi nó được đánh chỉ mục và lưu trữ trong môđun *Kho chứa tài liệu được chú thích NE và từ khóa*. Người dùng có thể tìm kiếm các tài liệu liên quan đến NE và/hoặc từ khóa quan trọng thông qua môđun *Tìm kiếm dựa trên NE và từ khóa*.

Trong mô hình VSM dựa trên NE và từ khóa [3], mỗi bộ ba NE hoặc từ khóa đều được xem như là một thuật ngữ tổng quát trong các vectơ truy vấn và vectơ tài liệu. Vectơ truy vấn $q$ và vectơ của một tài liệu thứ $j$, $d_j$, lần lượt được thể hiện như sau: $\vec{q} = (w_{1,q}, w_{2,q}, ..., w_{t,q})$, $\vec{d}_j = (w_{1,j}, w_{2,j}, ..., w_{t,j})$, với $t$ là tổng số thuật ngữ tổng quát trong bộ lưu trữ. Độ tương tự giữa truy vấn $q$ và tài liệu $d_j$ được tính bởi công thức:

$$\text{sim}(d_j, q) = \frac{\vec{d}_j \bullet \vec{q}}{|\vec{d}_j| \times |\vec{q}|} = \frac{\sum_{i=1}^{t} w_{i,j} \times w_{i,q}}{\sqrt{\sum_{i=1}^{t} w_{i,j}^2} \times \sqrt{\sum_{i=1}^{t} w_{i,q}^2}} \quad \text{(BT 2-1)}$$

Trong đó, $w_{i,j}$ và $w_{i,q}$ là trọng số của thuật ngữ tổng quát $t_i$ trong tài liệu $d_j$ và truy vấn $q$ được tính theo phương pháp *tf.idf* như trong mô hình VSM cổ điển.

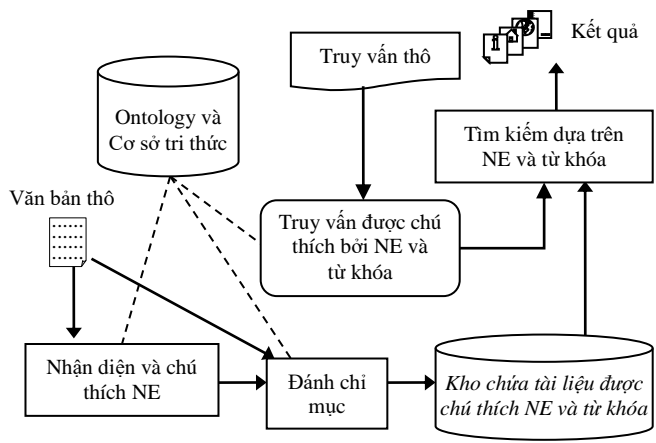

**Hình 1**. Truy hồi thông tin dựa trên thực thể có tên và từ khóa

Tuy nhiên, công trình [3] có các hạn chế sau đây:

- Hệ thống không biểu diễn được thông tin tìm ẩn của các từ để hỏi trong các truy vấn *Wh*, có mối liên hệ với các tài liệu mong đợi. Ví dụ như với truy vấn "*Where was George Washington born ?*", bên cạnh NE *George Washington* và từ khóa *born* thì từ để hỏi *Where* cũng chứa thông tin quan trọng, vì nó xác định câu trả lời được mong chờ là một nơi chốn nào đó.

- Với một NE có định danh thì NE này vẫn được biểu diễn theo dạng (*tên/lớp/định danh*). Do đó, để sử dụng tính chất bí danh của NE, [3] đã tiến hành mở rộng vectơ tài liệu bằng cách thêm vào các bí danh của NE này vào vectơ tài liệu có chứa NE này, tức là thêm (*tên khác/lớp/định danh*). Như vậy mỗi thực thể sẽ được gắn thêm nhiều bộ ba tương ứng với các tên khác nhau của nó, làm tốn bộ nhớ và thời gian xử lý hơn. Trong khi đó, với mô hình NEK+Wh chúng tôi đề xuất trong bài này nếu một NE có định danh thì chỉ biểu diễn nó bởi định danh *(*/*/định danh)*. Với cách biểu diễn như vậy, các sự xuất hiện của cùng một NE dưới các bí danh khác nhau vẫn so trùng được theo định danh của thực thể.

- Công trình [3] làm thực nghiệm trên tập kiểm tra TIME tương đối nhỏ, chỉ với 425 tài liệu và 83 truy vấn. Trong bài này chúng tôi đã sử dụng tập kiểm tra lớn hơn là tập L.A. Times, gồm 130.000 tài liệu và 124 truy vấn thuộc QA-Track-1999 của TREC.

## III. XỬ LÝ TÀI LIỆU VÀ TRUY VẤN

## IV. THỰC NGHIỆM

## V. CÔNG TRÌNH LIÊN QUAN

**1. Mở rộng vectơ tài liệu bằng khám phá mối quan hệ tiềm ẩn**

Trong các tài liệu, như là các trang web trên một website hoặc blog của một nhóm người nào đó, các NE, các từ khóa thường xuất hiện cùng nhau trong cùng một tài liệu. Ta nói các NE (hoặc các từ khóa) này có mối quan hệ đồng xuất hiện. [8] đề xuất việc tăng sự mô tả nội dung của một tài liệu bằng cách thêm các NE và từ khóa không xuất hiện trong tài liệu nhưng có mối quan hệ đồng xuất hiện với các NE và từ khóa đã tồn tại trong tài liệu. Bằng việc làm như vậy, chúng ta có thể bổ sung thêm các sự kiện không có trong tài liệu nhưng rất liên quan đến tài liệu. Ví dụ Zhu và Miller được xác định là có mối quan hệ đồng xuất hiện, nhưng trong tài liệu B chỉ đề cập đến Miller làm việc về W3C. Do đó, chúng ta có thể thêm Zhu vào tài liệu B để tăng thêm thông tin.

Mối quan hệ đồng xuất hiện được đo bằng độ mạnh quan hệ (relation strength). Có nhiều phương pháp tính độ mạnh quan hệ như Phi-squared, MI, VMI, Z score ([8]) và LRD (Latent Relation Discovery) ([16]). [8] đề xuất dùng phương pháp LRD để tính độ mạnh quan hệ, và mở rộng mở rộng vectơ tài liệu trong VSM (Vector Space Model) nhằm cải tiến việc truy hồi thông tin. Với việc mở rộng vectơ tài liệu, tác giả đã đạt được kết quả nhất định khi thực nghiệm trên 20 câu truy vấn ngẫu nhiên trong 112 truy vấn của tập CISI ([2]). Độ đo trung bình *F* tăng 19.3% so với 9.2% khi không mở rộng vectơ tài liệu. Đây là một kết quả tốt, tuy nhiên giá trị của độ *F* thấp, có thể là do tập kiểm tra CISI chứa rất ít thực thể có tên. Tuy nhiên công trình [8] còn có hạn chế như là: chỉ xem xét khía cạnh tên của NE, bỏ qua các đặc trưng khác của NE. Ví dụ như "Sài Gòn" có thể là sông Sài Gòn, đại học Sài Gòn, thành phố Sài Gòn hay thành phố Hồ Chí Minh. Do đó, hệ thống này không thể hỗ trợ các truy vấn có các dạng khác của NE như là theo định danh, theo cặp tên-lớp, theo lớp hoặc theo lớp bao phủ.

### 2. Mô hình không gian vectơ dựa trên Ontology

Hệ thống [5] nhận đầu vào là các truy vấn dạng RDQL (Resource Description Framework Data Query Language) chuẩn. Truy vấn RDQL tìm kiếm các NE thỏa mãn các điều kiện như thuộc về một lớp nào đó trong ontology và có các thuộc tính cụ thể (như là tác giả, ngày, nhà xuất bản,...). [5] đã mô hình hóa theo VSM truyền thống với các thành phần của không gian vectơ là các NE được xác định thông qua RDQL. Kết quả tìm được theo NE này được kết hợp với VSM theo từ khóa để có được tập kết quả cuối cùng trả về cho người dùng. Hướng tiếp cận này tương đối gần với hướng tiếp cận của chúng tôi nhất. Nhờ vào việc khai thác ontology và cơ sở tri thức, [5] đã giúp làm tăng độ đầy đủ khi tìm kiếm các NE, và tăng độ chính xác khi sử dụng các truy vấn có cấu trúc ngữ nghĩa theo dạng RDQL. Tuy nhiên, hệ thống còn nhiều khác biệt như:

- Quá trình kết hợp giữa NE và từ khóa để có tập kết quả cuối cùng là tuyến tính, với các trọng số là 0.5 bằng nhau lần lượt cho độ tương tự theo NE và từ khóa. Trong khi đó, mô hình VSM của chúng tôi đề xuất là mỗi thành phần của không gian vectơ là một thuật ngữ tổng quát, thuật ngữ này có thể là từ khóa, có thể là NE.

- Việc chuyển đổi truy vấn từ phía người dùng thành truy vấn các thực thể trong KB theo RDQL là một bước trung gian không cần thiết. Ví dụ, với truy vấn muốn tìm các tài liệu về các cầu thủ bóng chày, vectơ truy vấn theo NE được xác định với các thành phần là các cầu thủ bóng chày được tìm thấy trong KB. Điều này sẽ làm chậm tốc độ truy hồi của hệ thống. Trong mô hình NEK+Wh của chúng tôi, vectơ truy vấn và vectơ tài liệu đều đã có phần tử là lớp "*cầu thủ bóng chày*" và được so trùng vectơ bất cứ khi nào cần thiết.

- Sẽ dẫn đến việc bùng nổ vectơ truy vấn theo NE khi tìm kiếm các NE thuộc một lớp có nhiều NE thành viên trong Ontology qua truy vấn RDQL, ví dụ như là lớp *thành phố*, lớp *con người*,...

- Chỉ sử dụng tên và lớp của NE, không quan tâm đến bí danh của NE. Điều này sẽ dẫn đến sự thất bại là khi ta tìm các tài liệu về "thành phố Sài Gòn" thì các tài liệu về "thành phố Hồ Chí Minh" sẽ không được hệ thống truy hồi.

### 3. Các công trình liên quan khác

[15] sử dụng WordNet[3], là một ontology về ngôn ngữ, để mở rộng truy vấn. Công trình này chỉ dừng lại ở cấp độ từ, tức là ứng dụng các tập đồng nghĩa, tập gần nghĩa của từ khóa để làm gia tăng thêm thông tin cho truy vấn. Guha et al. [9] đề xuất các khái niệm phải được xác định trong truy vấn trước khi tìm kiếm, để có thể đạt được các trả lời liên quan. Ví dụ, nếu "*Washington*" có trong một truy vấn thì phải xác định rõ là nó nói về một người hoặc một trường đại học. Tuy nhiên, công việc này chủ yếu là làm tăng thêm dữ liệu để tìm kiếm kết quả.

Công trình [11], [6] chỉ ứng dụng lớp của NE kết hợp với từ khóa để gia tăng hiệu suất cho hệ thống, chưa xem xét hết các đặc tính của NE. Mặc khác, [6] chỉ khảo sát 2 lớp của NE đó là lớp *Phone* là lớp *Email*, và hệ thống chỉ hỗ trợ dạng truy vấn bắt buộc gồm lớp NE và từ khóa. Bên cạnh đó, ở công trình [13], chúng tôi đã đề xuất mô hình vectơ đa không gian khảo sát tất cả các đặc tính của NE như tìm kiếm theo lớp, theo tên, theo cặp tên, lớp hoặc theo định danh của NE. Tuy nhiên [13] chỉ khảo sát trên NE, chưa đề cập đến việc kết hợp NE và từ khóa.

## VI. KẾT LUẬN

Chúng tôi đã đề xuất mô hình không gian vectơ tổng quát hợp nhất thực thể có tên và từ khóa. Trong mô hình này, chúng tôi đã khai thác đầy đủ ba tính chất: tên, lớp, định danh của NE. Sử dụng đặc tính khác của NE là bí danh, lớp cha nhằm gia tăng sự thể hiện của tài liệu mà không làm thay đổi nội dung tài liệu này. Thêm vào đó, mô hình còn đáp ứng được nhiều loại truy vấn khác nhau như chỉ có NE, chỉ có từ khóa, có NE và từ khóa, kể cả các truy vấn dạng Wh-Question. Qua các kết quả thực nghiệm trên 130.000 tài liệu và 124 truy vấn, mô hình NEK+Wh đều có hiệu suất lớn hơn mô hình NEK và mô hình dựa trên từ khóa Lucene ở cả 2 độ đo được đề cập ở trên.

Độ tương tự, các mối quan hệ giữa từ khóa và thực thể có tên chưa được xem xét tới trong bài này. Đồng thời, việc khai thác độ tương tự giữa các thành phần của NE như độ tương tự về tên, lớp và cặp tên-lớp cũng chưa được khai thác. Nghiên cứu, khai thác các độ tương tự ngữ nghĩa này sẽ góp phần nâng cao hiệu suất truy hồi của hệ thống. Đây là một trong những hướng phát triển lớn của đề tài.

## TÀI LIỆU THAM KHẢO

---

[3] http://wordnet.princeton.edu/